\documentclass[aps, pre,reprint, a4paper, floatfix, superscriptaddress]{revtex4-1}
\usepackage{cmap}
\usepackage[T1]{fontenc}
\usepackage{times}

\pdfpageattr{/Group <</S /Transparency /I true /CS /DeviceRGB>>} 
\usepackage{newtxtext}
\usepackage[slantedGreek]{newtxmath}
\newcommand{\mathbold}[1]{\ensuremath{\textit{\textbf{#1}}}}
\DeclareSymbolFont{UPM}{U}{eur}{m}{n} 
\DeclareMathSymbol{\partial}{0}{UPM}{"40}

\newcommand{\SIseczref}[1]{SI~Sec.~\ref{#1}}
\newcommand{\SIfigzref}[1]{Fig.~\ref{#1}}

\newcommand{\refSub}[2]{\hyperref[#2]{\ref{#2}#1}}

\newcommand{\figrefsub}[2]{Fig.~\refSub{(#2)}{#1}}

\usepackage{helvet}
\usepackage{graphicx}
\usepackage{amsmath}
\usepackage{geometry}
\usepackage{comment}
\usepackage[british]{babel}
\usepackage{siunitx}
\usepackage{booktabs}

\makeatletter
\def\@bibdataout@aps{
 \immediate\write\@bibdataout{
 @CONTROL{
   apsrev41Control, author="08",editor="1",pages="0",title="0",year="1"
 }}
 \if@filesw
  \immediate\write\@auxout{\string\citation{apsrev41Control}}
 \fi
}
\makeatother

\usepackage[pdftex, bookmarks, pdffitwindow=false, pdfstartview=FitH, pdfdisplaydoctitle, colorlinks, plainpages=false, pdftitle={Theoretical prediction of thermal polarisation},pdfauthor={Peter Wirnsberger, Christoph Dellago, Daan Frenkel, Aleks Reinhardt}, pdfpagelabels, hypertexnames, citecolor={blue},linkcolor={red}, urlcolor={violet}, pdflang={en}, hyperfootnotes=false, breaklinks]{hyperref}
\usepackage[all]{hypcap}

\usepackage[dvipsnames]{xcolor}

\usepackage[stretch=17,shrink=17,step=1]{microtype}
\SetProtrusion{encoding={*},family={*},series={*},size={6,7}}
              {1={ ,550},2={ ,300},3={ ,300},4={ ,300},5={ ,300},
               6={ ,300},7={ ,400},8={ ,300},9={ ,300},0={ ,300}}

\usepackage{mathtools}
\usepackage{mleftright}

\newcommand{\pdc}[3]{%
  \mathchoice
  {\ensuremath{\left( \frac{\partial #1}{\partial #2}\right)_{#3}}}
  {\ensuremath{\left(\partial #1 / \partial #2 \right)_{#3}}}
  {\ensuremath{\left(\partial #1 / \partial #2 \right)_{#3}}}
  {\ensuremath{\left(\partial #1 / \partial #2 \right)_{#3}}}
}
\newcommand{\der}{\ensuremath{\mathrm{d}}}
\newcommand{\deriv}[2]{%
  \mathchoice
  {\ensuremath{\frac{\mathrm{d}#1}{\mathrm{d}#2}}}
  {\ensuremath{\mathrm{d} #1 / \mathrm{d} #2}}
  {\ensuremath{\mathrm{d} #1 / \mathrm{d} #2}}
  {\ensuremath{\mathrm{d} #1 / \mathrm{d} #2}}
}

\newcommand{\redDipole}{\ensuremath{p}}
\newcommand{\redTemp}{\ensuremath{T}}
\newcommand{\redRho}{\ensuremath{\rho}}

\begin{document}
\title[]{Theoretical prediction of thermal polarisation}
\author{P.~Wirnsberger}
\affiliation{Department of Chemistry, University of Cambridge, Lensfield Road, Cambridge, CB2 1EW, United Kingdom}
\author{C.~Dellago}
\affiliation{Faculty of Physics, University of Vienna, Boltzmanngasse 5, 1090 Wien, Austria}
\affiliation{Erwin Schr\"odinger Institute for Mathematics and Physics, University of Vienna, Boltzmanngasse 9, 1090 Wien, Austria}
\author{D.~Frenkel}
\affiliation{Department of Chemistry, University of Cambridge, Lensfield Road, Cambridge, CB2 1EW, United Kingdom}
\author{A.~Reinhardt}
\affiliation{Department of Chemistry, University of Cambridge, Lensfield Road, Cambridge, CB2 1EW, United Kingdom}

\begin{abstract}
We present a mean-field theory to explain the thermo-orientation effect in an off-centre Stockmayer fluid.
This effect is the underlying cause of thermally induced polarisation and thermally induced monopoles, which have recently been predicted theoretically. 
Unlike previous theories that are based either on phenomenological equations or on scaling arguments, our approach does not require any fitting parameters. 
Given an equation of state and assuming local equilibrium, we construct an effective Hamiltonian for the computation of local Boltzmann averages.
This simple theoretical treatment predicts molecular orientations that are in very good agreement with simulation results for the range of dipole strengths investigated. 
By decomposing the overall alignment into contributions from the temperature and density gradients, we shed further light on how the non-equilibrium result arises from the competition between the two gradients.
\end{abstract}

\maketitle

Temperature gradients are responsible for many fascinating coupling effects in condensed-matter systems,
such as driving flow via thermophoresis leading to mass separation in mixtures~\cite{Groot1984}, inducing thermo-osmotic slip along a surface~\cite{Ganti2017}, or causing thermoelectric effects through ion transport in electrolytes~\cite{Wurger2008, *Majee2013}.
In polar liquids, temperature gradients can induce appreciable electric fields even in the absence of ions by affecting the molecules' orientational order~\cite{Bresme2008}.
This `thermo-polarisation' (TP) effect is closely related to the thermo-orientation effect in non-polar fluids~\cite{Romer2012, Lee2016}.
The existence of the TP effect has been verified numerically for a range of different models of water~\cite{Bresme2008, Muscatello2011a, *Armstrong2013, *Iriarte-Carretero2016, Wirnsberger2016},
polar dumbbell molecules~\cite{Romer2012, Daub2016},
and an off-centre Stockmayer model for which the induced TP field is equivalent to one generated by a Coulomb charge~\cite{Wirnsberger2017}.

Microscopic interactions in systems out of thermal equilibrium are typically limited to a phenomenological treatment based on the theory of non-equilibrium thermodynamics~\cite{Groot1984}.
While this theory predicts a linear relationship between the TP field and the temperature gradient~\cite{Bresme2008}, it involves phenomenological coefficients that are unknown and require fitting to simulation data.
Recently, this shortcoming was addressed by using a mean-field theoretical approach~\cite{Lee2016},
which captured the scaling of the induced alignment of size-asymmetric polar dumbbell molecules accurately.
However, this approach entails an unknown parameter that is either estimated from dimensional arguments or fitted to simulation data.

\begin{figure}[t!]
\centering
\includegraphics{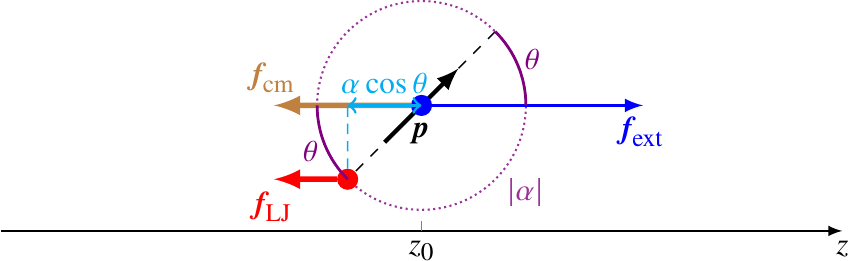}
\vspace{-\baselineskip}
\caption{\label{fig:sketch} 
Off-centre Stockmayer particle and its steady-state force balance.
The particle's centre of mass is located at $z_0$ and coincides with the location of the point dipole $\mathbold{p}$.
At steady state, the external force $\mathbold{f}_\text{ext}$ is balanced by the force $\mathbold{f}_{\text{cm\protect\vphantom{t}}}$ acting at the centre of mass and the force $\mathbold{f}_\text{LJ}$ acting at the LJ site.
Upon rotating the dipole by an angle $\theta$ relative to the $z$ axis, the $z$ position of the LJ site is $z_0+\alpha \cos\theta$ (since $\alpha < 0$).}
\end{figure}

In principle, it should be possible to eliminate this restriction and base quantitative predictions solely on the knowledge of the molecular geometry, the local thermodynamic state and the prevailing temperature and density gradients.
In this Letter, we propose a mean-field approach to predict the thermally induced alignment of an off-centre Stockmayer liquid without the need for any fitting parameters.
In this particularly simple model of a polar fluid~\cite{Wirnsberger2017}, the original Stockmayer potential~\cite{Stockmayer1941} is modified by displacing the Lennard-Jones (LJ) site from the position of the dipole $\mathbold{p}$, which continues to be located at the centre of mass.
Such asymmetry is required to produce forces that reorientate molecules in thermal gradients.
The relative displacement is given by $\alpha\hat{\mathbold{p}}$ (Fig.~\ref{fig:sketch}), where $\alpha$ is a parameter that controls the level of asymmetry, and $\hat{\mathbold{p}}=\mathbold{p}/|\mathbold{p}|$~\cite{Wirnsberger2017}.
In all cases studied here, we use $\alpha=-1/4$ (in reduced units; see \SIseczref{sec:reduced-units}), in line with previous work~\cite{Wirnsberger2017}.

To simplify the theoretical treatment, we assume that
(i)~the effect can be described by considering a single particle only,
(ii)~the average net force acting on this particle vanishes at steady state, and
(iii)~the system is at local equilibrium.
Suppose thermodynamic quantities and forces of interest vary along the $z$ axis.
The orientation of the particle can then be defined by the polar angle between its dipole moment vector and this axis,
$\cos\theta = \hat{\mathbold{p}} \cdot \hat{\mathbold{z}}$.
If we have an effective Hamiltonian $H(\cos\theta;\,z_0)$ that can account for the energy change upon rotating a particle about its centre of mass at position $z_0$ (Fig.~\ref{fig:sketch}), we can compute the Boltzmann-weighted average of the induced orientation~\cite{Lee2016},
\begin{equation}
\label{eq:boltzavg}
\langle \cos \theta(z_0) \rangle = \frac{1}{Q(z_0)} \int_{-1}^{1} q \exp[-\beta(z_0) H(q;\,z_0)] \, \der q,
\end{equation}
where $Q(z_0) = \int_{-1}^{1} \exp[-\beta H(q;\,z_0)]\,\der q$, $\beta(z_0) = 1/k_\text{B}T(z_0)$, $T(z_0)$ is the absolute temperature at $z_0$ and $k_\text{B}$ is Boltzmann's constant.
Integration over the azimuth angle is omitted because contributions in the numerator and the denominator cancel out.

According to assumption~(ii), on average all forces acting on a particle add up to zero so that the system does not undergo continual net acceleration.
Individual forces can, however, act at different locations within the particle and thereby generate a torque.
There are three possible sites at which these forces could attach in our model system: the LJ site, the point dipole site and the centre of mass.
Since the latter two coincide, any external force $f_\text{ext} \hat{\mathbold z}$ acting on the particle must be balanced by the sum of a force $f_{\text{cm\vphantom{t}}} \hat{\mathbold z}$ acting on the centre of mass and a force $f_\text{LJ} \hat{\mathbold z}$ acting on the LJ site, so that
\begin{equation}
\label{eq:forcebalance}
f_\text{LJ} + f_{\text{cm\vphantom{t}}} + f_\text{ext} = 0.
\end{equation}
The overall torque acting on the particle is therefore given by $\alpha \hat{\mathbold{p}}\times \mathbold{f}_\text{LJ}$.
An infinitesimal rotation changes the energy by $-f_\text{LJ} \alpha \, \der(\cos\theta)$.
However, this rotation polarises the liquid, and we must therefore account for the coupling between the dipole and the mean electric field $\langle E(z_0)\rangle$.
Combining both contributions, the work done to rotate the particle is approximated as
\begin{equation}
\label{eq:Hamiltonian}
H(\cos\theta; z_0) =  -\alpha \cos\theta f_\text{LJ}(z_0) - p \cos\theta \langle E(z_0) \rangle.
\end{equation}
Using this Hamiltonian, Eq.~\eqref{eq:boltzavg} yields $\langle\cos\theta(z_0)\rangle  = \coth\xi - \xi^{-1} = \xi/3 + O(\xi^3)$,
where $\xi = \beta(z_0)\alpha f_\text{LJ}(z_0) + \beta(z_0) p \langle E(z_0) \rangle$.
For $|\xi|\ll 1$, we can truncate the series at linear order in $\xi$.
We further approximate the field self-consistently using~\cite{Lee2016}
\begin{equation}
\label{eq:eselfcon}
\langle E(z_0) \rangle  \approx -4\uppi \rho(z_0) p \langle \cos\theta(z_0) \rangle,
\end{equation}
where $\rho(z_0)$ is the number density, and thus obtain our central result for the mean alignment,
\begin{equation}
\label{eq:costheory}
\langle \cos \theta(z) \rangle = \frac{\beta(z) \alpha f_\text{LJ}(z)}{3+4 \uppi \rho(z) \beta(z) p^2}.
\end{equation}
Apart from the nature of the force, this expression is analogous to Eq.~(11) of Ref.~\citenum{Lee2016}.

\begin{figure}
\centering
\includegraphics{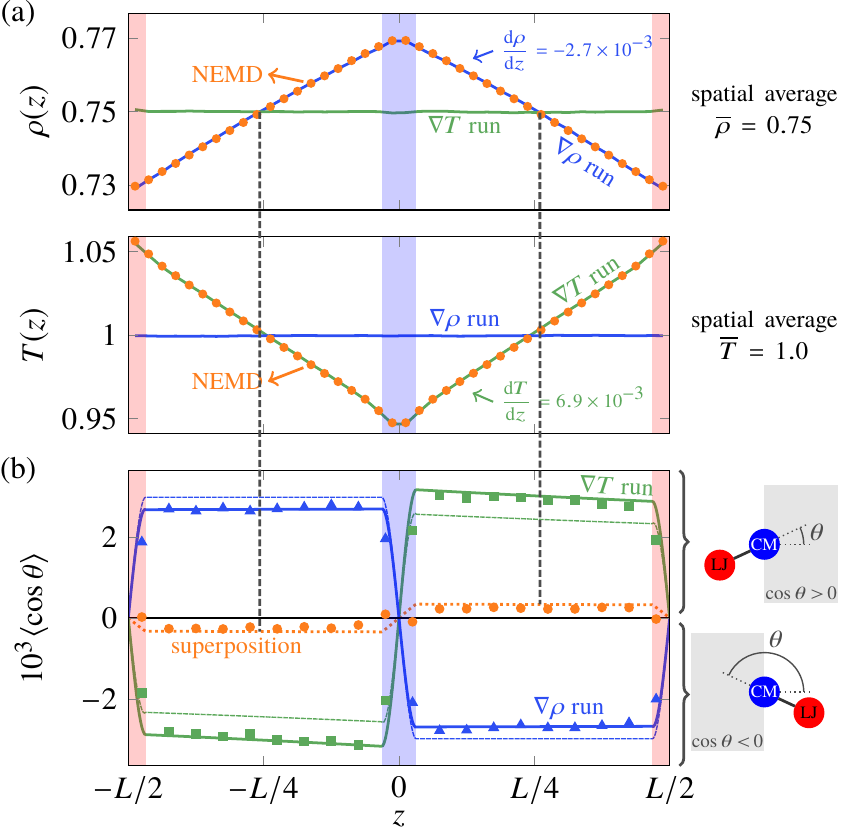}
\vspace{-1.5\baselineskip}
\caption{\label{fig:comb-trho-costheta-LJ}
Off-centre LJ. $L=31.45$.
(a)~Density and temperature profiles for NEMD, $\nabla\rho$ and $\nabla T$ runs with reservoirs shaded.
(b)~Mean orientation from simulation (symbols) and theoretical predictions (solid lines).
Dashed coloured lines correspond to incorrect attachment sites of $f_\text{ext}^\text{id}$.
Points at which force superposition (dotted line) is expected to perform well are indicated by dashed vertical lines.
Symbols are larger than error bars.}
\end{figure}

To benchmark the predictions of this equation against simulation results, we ran non-equilibrium molecular dynamics (NEMD) simulations of an off-centre Stockmayer fluid in a quasi-one-dimensional geometry.
We imposed temperature and density gradients by simulating a hot and a cold reservoir orthogonal to the $z$ axis~\cite{Wirnsberger2016}, so that all thermodynamic driving forces varied only along $z$.
Following the approach of Daub~\textit{et al.}~\cite{Daub2016}, we disentangled the effects of the two gradients by applying a body force to each particle.
In `$\nabla\rho$ runs', we generated a density gradient by applying a body force to the centre of mass of each particle in an equilibrium $NVT$ simulation.
Similarly, in `$\nabla T$ runs', we eliminated the density gradient by applying a body force of opposite sign to the NEMD run.
This procedure allowed us to decompose the non-equilibrium phenomenon into an equilibrium problem at constant temperature and a non-equilibrium problem at constant density.
We show typical $\rho$ and $T$ profiles of all three runs for an off-centre LJ system (i.e.~at zero dipole strength) in \figrefsub{fig:comb-trho-costheta-LJ}{a}.
We adjusted the applied body forces to yield gradients within \SI{1}{\percent} of the NEMD results.
Full technical details of the simulation set-up are given in \SIseczref{sec:sim-details}.

The applied body force in simulation corresponds to $f_\text{ext}$ in Fig.~\ref{fig:sketch}. 
To determine $f_\text{LJ}$, we first compute $f_{\text{cm\vphantom{t}}}$ and then invoke the force-balance argument [Eq.~\eqref{eq:forcebalance}], $f_\text{LJ} = -f_\text{ext} -f_{\text{cm\vphantom{t}}}$.
In the absence of a dipole, the centre of mass itself does not exhibit any direct interaction with other particles.
In simulations, forces are computed from gradients of the potential energy, and so do not include thermodynamic forces arising from momentum degrees of freedom.
The $z$ component of the average force derived from the potential energy $\langle f_i \rangle$ balances such `ideal' forces at steady state;
we demonstrate in \SIseczref{sec:idealforce} that $\langle f_i\rangle =  f_\text{ext}^\text{id}$, where
\begin{equation}
  \label{eq:idealforce}
  f_\text{ext}^\text{id} = \begin{cases}  \left(\frac{\partial \mu^\text{id}}{\partial \rho}\right)_T \frac{\der \rho}{\der z} = \frac{k_\text{B} T}{\rho} \frac{\der \rho}{\der z} & \text{ for $\nabla\rho$ runs,} \\[2mm]
   \left[s^\text{id} + \left(\frac{\partial \mu^\text{id}}{\partial T}\right)_{\rho}\right]\frac{\der T}{\der z} = k_\text{B} \frac{\der T}{\der z}& \text{ for $\nabla T$ runs,}
    \end{cases}
\end{equation}
with $\mu^\text{id}$ being the ideal chemical potential and $s^\text{id}$ the ideal entropy per particle.

To compute $f_\text{LJ}$, we need to consider which part of the ideal force exerts a torque.
Although this point is not immediately obvious, comparison with simulation data for $\langle \cos \theta \rangle$ answers this question unambiguously:
the term proportional to $\nabla\rho$ does not generate a torque, and we can interpret it as a force acting on the centre of mass ($f_{\text{cm\vphantom{t}}}^{(\nabla\rho)}= -f_\text{ext}^\text{id}$).
The term proportional to $\nabla T$, however, generates a torque that we can translate into an apparent force acting on the LJ site (so that $f_{\text{cm\vphantom{t}}}^{(\nabla T)}=0$).
We find it somewhat curious that the two gradients behave differently in this respect; however, we do not at present have a clear physical argument for this behaviour.
The force-balance argument~[Eq.~\eqref{eq:forcebalance}] thus gives
\begin{equation}
  \label{eq:flj}
     f_\text{LJ} = -f_\text{ext} + 
     \begin{cases}  
        \frac{k_\text{B} T}{\rho} \frac{\der \rho}{\der z}    &\text{for $\nabla\rho$ runs,} \\
       0                                                         &\text{for $\nabla T$ runs.}
     \end{cases}
\end{equation}
Using this expression and fits to the sampled density and temperature profiles, we employ Eq.~\eqref{eq:costheory} to obtain a theoretical estimate of $\langle \cos\theta\rangle$.

We show the theoretical prediction and simulation data for a representative thermodynamic state in \figrefsub{fig:comb-trho-costheta-LJ}{b}, alongside alternative  predictions with the apparent attachment sites of $f_\text{ext}^\text{id}$ swapped (dashed lines), demonstrating that our choice of attachment sites in Eq.~\eqref{eq:flj} is correct.
Theoretical predictions agree remarkably well with simulation data in the regions outside the reservoirs.
Close to the reservoirs, the agreement is worse, perhaps because of the discontinuous nature of the employed thermostat, which creates an interface and likely violates our local equilibrium assumption.
Interestingly, the density gradient induces a rotation of the LJ site towards lower densities, while the temperature gradient has the opposite effect.
In each case, the alignment is an order of magnitude larger than in the NEMD case.
These individual gradients also allow us to estimate the NEMD result for the points at which $\rho$ and $T$ agree in all three runs, i.e.~at $z = \pm L/4$, where $\rho=\overline{\rho}$ and $T=\overline{T}$.
To this end, we summed the two forces in Eq.~\eqref{eq:flj} and employed Eq.~\eqref{eq:costheory} together with the NEMD density and temperature profiles.
We show in \figrefsub{fig:comb-trho-costheta-LJ}{b} that the superposition estimate is in rather good agreement with the simulation result not only for $z=\pm L/4$, but everywhere outside the reservoirs.
The good agreement with simulation results demonstrates that our theory captures the underlying physics accurately in the non-polar case and suggests that the assumptions underlying our theoretical treatment are reasonable.

Next, we studied the behaviour of an off-centre Stockmayer fluid for different dipole strengths ${\redDipole}^2\in[0,3]$.
To make comparisons meaningful, we chose a reference state of $\overline{\redTemp}=1.25$ and $\overline{\redRho}=0.82$, at which the equilibrium Stockmayer fluid is a liquid for all dipole strengths considered~\cite{Kriebel1996}.
We imposed the same temperature gradient in all NEMD runs by fixing the reservoir temperatures while letting the induced density vary (\SIfigzref{fig:comb-rhoT-stock}).
To rule out the presence of nematic or ferrofluidic phases, we measured the correlation functions $h_{110}(r)$ and $h_{220}(r)$ and verified that both vanish as $r\to\infty$~\cite{Weis1993}.
As in the non-polar case, the crucial quantity for our theory is $f_\text{LJ}$, whose estimation is now complicated by dipole--dipole interactions.
The dipole is located at the centre of mass, and we assume any dipole-induced isotropic force contribution attaches to that site, adding to $f_{\text{cm\vphantom{t}}}$.

Suppose $f_\text{ext}(p^2)$ is the external force that generates the target density gradient in a $\nabla\rho$ run with dipole strength $p^2$.
This force differs from that required to generate the same density gradient in the non-polar case by
$  \upDelta f_\text{dip}(p^2) = f_\text{ext}(p^2) - f_\text{ext}(0)$.
Since all other parameters are kept fixed, we can attribute the entirety of this force to the presence of the dipole.
An analogous result holds for the $\nabla T$ run, where the force difference is evaluated for a fixed temperature gradient at constant density.
In order to employ the force-balance argument [Eq.~\eqref{eq:forcebalance}] to compute $f_\text{LJ}$, we also need to determine where the ideal forces [Eq.~\eqref{eq:idealforce}] effectively attach.
In the non-polar case, the ideal force proportional to $\nabla\rho$ acted at the centre of mass.
We do not expect this behaviour to change for ${\redDipole}^2\ne0$ because the dipole is located at the same site.
However, the situation is less clear for the $\nabla T$ contribution, which we initially assigned solely to the LJ site.
It is not unreasonable to assume that for sufficiently large ${\redDipole}^2$, because the dipole is at the centre of mass, some part of this balancing ideal force will also act at this site, but the dipole strength at which this shift becomes relevant to our theory is not obvious.
For a Stockmayer system, a dipole strength of ${\redDipole}^2 = 0.25$ constitutes an almost negligible perturbation to pure LJ~\cite{Stell1972}, and we do not therefore expect the behaviour to change qualitatively in this case.
However, electrostatic contributions increase rapidly with ${\redDipole}^2$~\cite{Stell1972}; thus, for ${\redDipole}^2 \ge 1$, we split the ideal balancing force equally between $f_\text{LJ}$ and $f_\text{cm\vphantom{t}}$, i.e.~the ideal $\nabla T$ force is assumed to generate only half the torque compared to the ${\redDipole}^2 < 1$ case.
As we lack further insight into the precise mechanism by which this process occurs, our choice is rather arbitrary; for completeness, we provide the results for an alternative choice in \SIfigzref{fig:comb-costheta-stock-altIdeal}.
Gathering all contributions and assuming force balance [Eq.~\eqref{eq:forcebalance}], we find
\begin{equation}
  \label{eq:fljstock}
     f_\text{LJ}(p^2) = -f_\text{ext}(0) +
     \begin{cases}  
        \frac{k_\text{B} T}{\rho} \frac{\der \rho}{\der z}    &\text{for $\nabla\rho$ runs,} \\
       0                                                      &\text{for $\nabla T$ runs if ${\redDipole}^2 < 1$,} \\
       \frac{1}{2} k_\text{B} \frac{\der T}{\der z}          &\text{for $\nabla T$ runs if ${\redDipole}^2 \ge 1$.}
     \end{cases}
\end{equation}
Apart from the behaviour of the ideal force, this result is identical to Eq.~\eqref{eq:flj}.
The difference between $f_\text{LJ}(p^2)$ and $f_\text{LJ}(0) \equiv f_\text{LJ}$ is that all terms are evaluated for a different thermodynamic state, since the density profile changes with dipole strength (\SIfigzref{fig:comb-rhoT-stock}).
\begin{figure}[b!]
\centering
\includegraphics{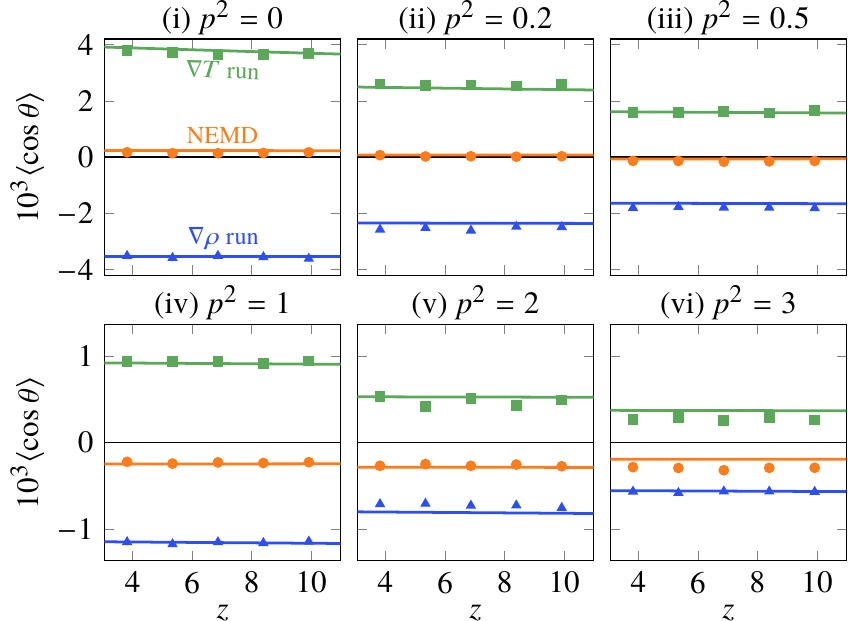}
\caption{\label{fig:comb-costheta-stock}
Mean orientation from simulation (symbols) and theoretical predictions (solid lines) for an off-centre Stockmayer liquid at varying dipole strength. $L=30.53$.
Only the region between the reservoirs is shown for compactness.
Results were averaged over both halves of the simulation box to improve statistics.}
\end{figure}

We compare simulation results to the theoretical estimates obtained with Eqs~\eqref{eq:costheory} and~\eqref{eq:fljstock} (Fig.~\ref{fig:comb-costheta-stock}). 
For the $\nabla\rho$ and $\nabla T$ runs, the magnitude of the alignment is maximal for $\redDipole=0$ and decreases quickly with $\redDipole$ due to the energetically unfavourable interaction with the total electric field.
For the highest dipole strength (${\redDipole}^2 = 3$), the induced alignment is approximately an order of magnitude lower than in the non-polar case.
Intriguingly, the presence of the dipole flips the sign of the orientation in the NEMD run, a feature not observed for size-asymmetric polar dumbbell molecules~\cite{Daub2016}.
The crossover occurs close to ${\redDipole}^2 = 0.2$, where the alignments from the density and temperature gradients almost exactly compensate one another.
Furthermore, the NEMD result remains largely unchanged for ${\redDipole}^2 \ge 1$, perhaps indicating that the effect becomes saturated.
Overall, our theoretical prediction agrees very well with the simulation data.
By construction, absolute errors in the predictions for the individual gradient contributions propagate to the prediction for the NEMD result because we add up the two forces.
This point is well illustrated for ${\redDipole}^2= 3$, where the $\sim$\SI{20}{\percent} error in the prediction of the $\nabla T$-run result causes the estimate for the NEMD run to be shifted by the same constant.
Despite this limitation, these results suggest our mean-field theory captures the essential physics underlying this non-equilibrium phenomenon very well.

So far, we have treated $f_\text{LJ}$ as an input parameter and determined it numerically to match a density gradient.
However, we may be able to predict the force acting on the LJ site from the equation of state (EOS), for which accurate fits exist~\cite{Johnson1993,Thol2016}.
Starting from the Gibbs--Duhem relation, we find an explicit expression for the external force in terms of the chemical potential $\mu$ (\SIseczref{sec:force}),
\begin{equation}
  f_\text{ext} = \pdc{\mu}{\rho}{T} \deriv{\rho}{z} + \left[s + \pdc{\mu}{T}{\rho}\right]\deriv{T}{z}. \label{eq:theory-fext}
\end{equation}
At local equilibrium, this force will be exactly balanced by a pressure-gradient force
\begin{align}
 f_\text{balance} = f_{\text{cm\vphantom{t}}}+f_\text{LJ} = -\frac{1}{\rho}\left[ \pdc{P}{\rho}{T}\deriv{\rho}{z} + \pdc{P}{T}{\rho}\deriv{T}{z} \right],
\end{align}
so that the sum of both forces vanishes.
For $\nabla\rho$ runs, all terms in Eq.~\eqref{eq:theory-fext} involving a temperature gradient vanish, and vice versa for $\nabla T$ runs, so the external force can be written as
\begin{equation}
  \label{eq:fljstock2}
     f_\text{ext}(0) =  
     \begin{cases}  
        \pdc{\mu}{\rho}{T}  (\deriv{\rho}{z})                 &\text{for $\nabla\rho$ runs,} \\
        \big[s + \pdc{\mu}{T}{\rho}\big](\deriv{T}{z})     &\text{for $\nabla T$ runs.} \\
     \end{cases}
\end{equation}
Therefore, to generate (say) a density gradient, instead of adjusting the external force until we get agreement with NEMD results, we can use the Johnson EOS~\cite{Johnson1993} and Eq.~(\ref{eq:fljstock2}) to \textit{predict} it by multiplying the derivative of the chemical potential by the density gradient we wish to match.
For suitably large system sizes, simulation results are in excellent agreement with the target value (deviation $<$\SI{1}{\percent}).
The predicted force cancelling the gradient in $\nabla T$ runs is less accurate (deviation $\sim$\SI{2.5}{\percent}).

\begin{figure}[t!]
\centering
\includegraphics{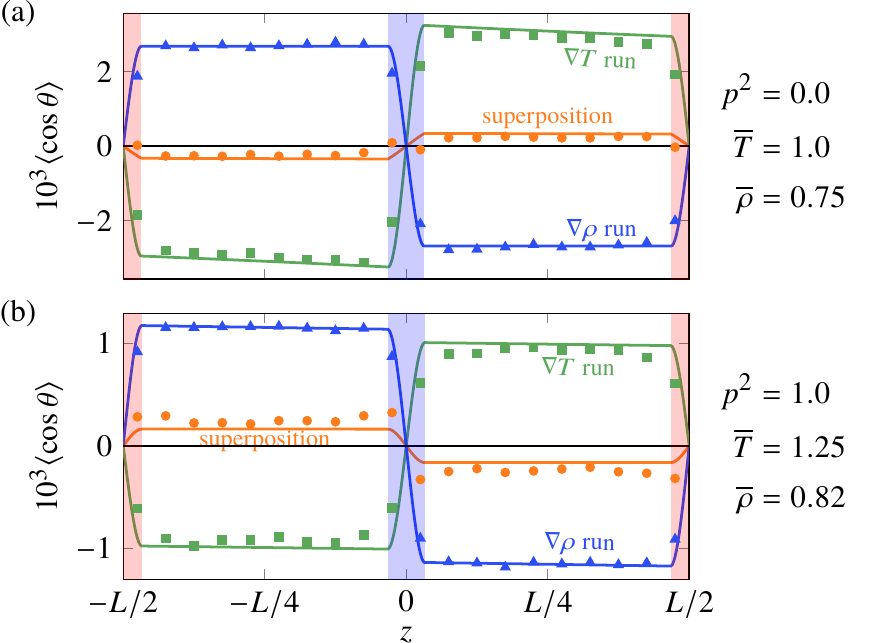}
\caption{\label{fig:comb-costheta-fullpred}
Mean orientation from simulation (symbols) and theoretical predictions (solid lines) using an EOS-derived estimate of $f_\text{LJ}$. 
Symbols are larger than error bars.
States corresponding to (a)~Fig.~\ref{fig:comb-trho-costheta-LJ} and (b)~\figrefsub{fig:comb-costheta-stock}{iv}.}
\end{figure}

A comparison of simulation results with the theoretical prediction of Eqs~\eqref{eq:costheory} and \eqref{eq:fljstock2} using the Johnson EOS~\cite{Johnson1993} is shown in Fig.~\ref{fig:comb-costheta-fullpred}. 
For off-centre LJ (\figrefsub{fig:comb-costheta-fullpred}{a}), the predictions are almost as accurate as the ones obtained with the numerically determined force (\figrefsub{fig:comb-trho-costheta-LJ}{b}).
The agreement is excellent for $\nabla\rho$ runs and exhibits only a slight deviation ($\sim$\SI{5}{\percent}) for $\nabla T$ runs.
Superposition of the forces yields again an accurate estimate for the NEMD result with a deviation of only $\sim$\SI{20}{\percent}.
For the polar case (\figrefsub{fig:comb-costheta-fullpred}{b}), the EOS estimate yields very accurate predictions for $\nabla\rho$ runs and overestimates the alignment in $\nabla T$ runs only slightly.
The prediction for NEMD runs is shifted by the same constant and is off by $\sim$\SI{30}{\percent} compared to the simulation result.
Although the estimates are not as accurate as the original ones (Fig.~\ref{fig:comb-costheta-stock}), we find it remarkable that such good agreement can be achieved solely based on the LJ EOS.
While not perfect, this route provides a particularly facile back-of-the-envelope estimate of the degree of molecular alignment without needing any simulations at all.
Interestingly, while our theory is approximate and relies on the assumption of local equilibrium, it gives reliable estimates even for the very large gradients considered here ($|\deriv{T}{z}|\sim \SI{2e9}{\kelvin\per\metre}$ for typical molecular LJ parameters).

In summary, we have proposed a mean-field theory to explain the thermo-orientation and thermo-polarisation effects exhibited by an off-centre Stockmayer liquid.
Our theoretical predictions are in very good agreement with simulation results for a range of dipole strengths, including in the absence of any dipole.
In line with previous work~\cite{Lee2016}, we found that the two effects are caused by the same underlying physical mechanism.
Differences in the predicted alignment as a function of the dipole strength are mainly caused by the energetic penalty for forming an electric field and the behaviour of ideal forces when dipoles are present.
By separating temperature and density gradients using applied body forces, we found the individual contributions lead to an alignment of opposite sign, which can be rationalised by the requirement for overall force balance in the non-equilibrium steady state.
Finally, we demonstrated that very reasonable predictions can be obtained solely based on the LJ EOS and the non-equilibrium density and temperature profiles.
In future work, it will be interesting to see whether our theory can be extended to water, where quadrupolar interactions are known to play an important role in thermal polarisation~\cite{Wirnsberger2016, Armstrong2015a}.

\begin{acknowledgements}
We acknowledge helpful discussions with Carl Poelking, Michiel Sprik and Robert Jack.
This work was supported by the Erwin Schr\"odinger Institute for Mathematics and Physics through a Junior Research Fellowship,
the Cambridge Philosophical Society through a Research Studentship
and the Austrian Science Fund (FWF) [SFB ViCoM, Project F41].
The results presented here were achieved in part using the Vienna Scientific Cluster.
Supporting data are available at the University of Cambridge Data Repository, \href{https://doi.org/10.17863/cam.22951}{doi:10.17863/cam.22951}~\cite{supporting-data}.
\end{acknowledgements}


\makeatletter
\renewcommand*{\thesection}{S\arabic{section}}
\renewcommand*{\thesubsection}{S\thesection.\arabic{subsection}}
\renewcommand*{\p@subsection}{}
\renewcommand{\p@subsubsection}{}
\renewcommand*{\theHfigure}{\thepart.\thefigure}
\renewcommand*{\theHequation}{\thepart.\theequation}
\renewcommand*{\theHsection}{\thepart.\thesection}
\renewcommand{\thefigure}{S\arabic{figure}}
\renewcommand{\theequation}{S\arabic{equation}}
\renewcommand\@seccntformat[1]{\csname the#1\endcsname\quad}
\makeatother

\setcounter{section}{0}
\setcounter{equation}{0}

\section*{Supplementary Information}
\vspace{\baselineskip}

\section{Reduced units}\label{sec:reduced-units}
We non-dimensionalise all quantities in our work in terms of $\varepsilon$, the Lennard-Jones (LJ) well depth; $\sigma$, the LJ diameter; $k_\text{B}$, Boltzmann's constant; $\varepsilon_0$, the electric constant; and $m$, the particle mass, as shown in the following table.
\begin{center}
\begin{tabular}{p{3cm} l}
\toprule
\textbf{Reduced quantity} & \textbf{Definition}\\
\midrule
distance $\ell^*$ & $\ell/\sigma$ \\
number density  $\rho^*$ & $\rho \sigma^3$ \\
force    $f^*$    & $f \sigma/\varepsilon$ \\
time     $t^*$    & $t/\sqrt{\sigma^2 m/\varepsilon}$\\
temperature $T^*$ &  $k_\text{B}T/\varepsilon$ \\
dipole moment $p^*$ & $p/\sqrt{4\uppi \varepsilon_0\varepsilon\sigma^3}$\\
\bottomrule
\end{tabular}
\end{center}
For notational simplicity, we have not denoted reduced quantities with asterisks in the main text.

\section{Simulation details}\label{sec:sim-details}
We used a modified version of the \textsc{Lammps} simulation package~\cite{Plimpton1995} (v.~11Aug17) to perform all molecular dynamics simulations presented in this work.
In all cases, we employed a fully periodic tetragonal simulation box with dimensions $3L_x = 3L_y = L_z \equiv L$ containing \num{5832} off-centre Stockmayer particles~\cite{Wirnsberger2017}.
The off-centre potential is a modification of the original Stockmayer pair potential~\cite{Stockmayer1941} for which the LJ site is displaced from the dipole site by a vector $\alpha \hat{ \mathbold{p}}$, where $\alpha$ is a control parameter that we set to $-\sigma/4$, and $\hat {\mathbold p}$ is the unit vector of the dipole moment (Fig.~\ref{fig:sketch}).

The off-centre Stockmayer potential is given by~\cite{Wirnsberger2017}
\begin{equation}
\label{eq:phi}
 \phi_{ij} = 4\varepsilon \left[ \left(\frac{\sigma}{\xi_{ij}}\right)^{12}-\left(\frac{\sigma}{\xi_{ij}}\right)^{6} \right] -
\frac{\mathbold{p}_i\cdot \mathbold{p}_j-3(\mathbold{p}_i\cdot \skew{1.5}\hat{\mathbold{r}}_{ij})(\skew{1.5}\hat{\mathbold{r}}_{ij} \cdot \mathbold{p}_j)}{4\uppi \varepsilon_0 r_{ij}^3},
\end{equation}
where $\mathbold{r}_i$ and $\mathbold{p}_i$ are the position vector and dipole moment vector of particle $i$,
$\mathbold{r}_{ij} = \mathbold{r}_j - \mathbold{r}_i$ is the interparticle vector,
$\hat{\mathbold{r}}_{ij} = \mathbold{r}_{ij}/r_{ij}$ and $\xi_{ij}=|\mathbold{r}_{ij} + \alpha (\hat{\mathbold{p}}_j - \hat{\mathbold{p}}_i)|$.
Electrostatic interactions are not affected by the displacement of the LJ site and were treated with Ewald summation and tinfoil boundary conditions~\cite{Ewald1921, *deLeeuw1980, *Toukmaji2000}.
We set the cutoff radius to $7\sigma$ for all types of interaction and specified a relative accuracy of approximately $10^{-5}$ for the computation of long-range forces.
\textsc{Lammps} source files containing our implementation of the off-centre Stockmayer potential are available for download with the supporting data.

To equilibrate our systems, we followed the protocol outlined in Ref.~\citenum{Wirnsberger2017}.
We first generated a lattice structure with random dipole orientations and equilibrated it in an $NVT$ simulation for at least $2\times 10^3 \tau$,
where $\tau = \sigma \sqrt{m/\varepsilon}$ is the unit of time.
In $NVT$ simulations, we set the relaxation time of the Nos\'e--Hoover thermostat~\cite{Nose1984, *Hoover1985} to $0.5 \tau$.
We employed a time step of $\upDelta t = 0.005 \tau$ for the discrete time integration
for the simulations corresponding to Fig.~\ref{fig:comb-trho-costheta-LJ} and $\upDelta t = 0.004\tau$ for the ones corresponding to Fig.~\ref{fig:comb-costheta-stock}.
In $\nabla\rho$ runs, we imposed a body force acting on each particle during this equilibration run in order to establish the desired density gradient.
Subsequently, we performed the production run retaining this body force, and we sampled spatial profiles for temperature, density and molecular orientation.
For $\nabla T$ runs and full NEMD runs, we adjusted velocities of the last configuration of the $NVT$ simulation so that the total energy matched the average energy sampled~\cite{Wirnsberger2015}.
We then equilibrated the system for another $\num{2e3}\tau$ in an $NVE$ simulation before imposing either a heat flux using the eHEX algorithm~\cite{Wirnsberger2015} or a temperature
gradient by applying two Gaussian thermostats locally to adjust the non-translational kinetic energy inside the reservoirs appropriately.
The former approach is suitable when imposing a constant heat flux (e.g.~the results presented in Fig.~\ref{fig:comb-trho-costheta-LJ}),
while the latter is better suited when a constant temperature gradient is desired (e.g.~the results presented in Fig.~\ref{fig:comb-costheta-stock}).

We waited for at least $\num{2e3}\tau$ for a steady state to become fully established and for any transient behaviour to vanish before starting production runs.
For $\nabla T$ runs, we imposed a body force both during the steady-state equilibration and in the production run in order to remove the density gradient.
Production runs were simulated for between $\num{1e5}\tau$ and $\num{5e5}\tau$;
we stopped simulations when the statistics for $\langle \cos \theta \rangle$ were sufficiently converged.
Even for the longest simulation (the NEMD run in Fig.~\ref{fig:comb-trho-costheta-LJ}), we did not observe any energy loss with the eHEX algorithm.
As pointed out in Ref.~\citenum{Wirnsberger2016}, the piecewise constant profile for $\langle \cos \theta \rangle$ is established fairly quickly, but because there is no energetic penalty for having a net dipole moment with tinfoil boundary conditions, long simulation times may be required for $\langle\cos\theta\rangle$ to be centred around zero perfectly.
Since the constant term $(1/L) \int_L \langle \cos\theta (z) \rangle\,\der z$ is very small and must vanish by symmetry for an infinitely long run, we are justified in subtracting it from the sampled profile $\langle \cos \theta (z) \rangle$ to reduce the computational cost.

Imposing a piecewise constant force proportional to $\operatorname{sgn}(z) \hat{\mathbold{z}}$ leads to a serious drift in the total energy in $NVE$ simulations.
We found this problem to be related to the discontinuity of the force at the origin when a piecewise constant force was applied.
The problem was only observable in $NVE$ simulations or in combination with the eHEX algorithm, because in all other cases the lost energy is re-supplied by the thermostat.
To resolve it, we fitted third-order polynomials inside the reservoirs so that the resulting force profile was continuously differentiable.
This procedure eliminated the energy drift completely.

\section{Force derivation}\label{sec:force}
In this section, we derive the analytical force expressions given by Eqs~(\ref{eq:idealforce})--(\ref{eq:fljstock}).
We start from the fundamental equation for the internal energy $E$,
\begin{equation}
 \der E = T\,\der S - P\,\der V + F\,\der z + \mu\,\der N,
\end{equation}
where $S$ is the entropy, $P$ the pressure, $V$ the volume, $F$ is a force acting on the system, $\mu$ the chemical potential and $N$ the number of particles.
The Gibbs energy is given by $G=E-TS+PV = \mu N$, so its total differential can be written as
\begin{equation}
\der G =  F\,\der z + \mu\,\der N  - S\,\der T + V\,\der P =  N\,\der \mu + \mu\,\der N.
\end{equation}
After division by $N$, we obtain the Gibbs--Duhem analogue
\begin{equation}
\label{eq:mugibbsduhem}
\der \mu = f\,\der z - s\,\der T + v\,\der P,
\end{equation}
where $v=1/\rho$.
Since $\mu=\mu(\rho,\,T)$ and $P = P(\rho,\,T)$,
we can write the total differentials of both functions as
\begin{align}
 \der \mu & = \pdc{\mu}{\rho}{T}\,\der \rho + \pdc{\mu}{T}{\rho}\,\der T \quad \text{and}\\ 
 \der P & = \pdc{P}{\rho}{T}\,\der \rho + \pdc{P}{T}{\rho}\, \der T.
\end{align}
Temperature and density vary only with $z$ and their total differentials are given by
\begin{equation}
\label{eq:derRho-derT}
 \der \rho = \deriv{\rho}{z}\,\der z  \qquad \text{and}\qquad \der T = \deriv{T}{z}\,\der z.
\end{equation}
Combining Eqs~\eqref{eq:mugibbsduhem}--\eqref{eq:derRho-derT} and comparing the coefficients of the differentials, we find that the force per particle can be expressed as
\begin{equation}
\begin{split}
 f &=  \underbrace{\pdc{\mu}{\rho}{T} \deriv{\rho}{z} + \left[s + \pdc{\mu}{T}{\rho}\right]\deriv{T}{z}}_{f_\text{ext}} \\
 & \qquad {}  \underbrace{-\frac{1}{\rho}\left[ \pdc{P}{\rho}{T}\deriv{\rho}{z} +  \pdc{P}{T}{\rho}\deriv{T}{z} \right]}_{f_\text{balance}}.
\end{split}\label{eq-fullF}
\end{equation}
The overall thermodynamic force acting on a particle vanishes at equilibrium, i.e.~$f=0$.
If we identify the force $f_\text{ext}$ as an externally applied force,
at equilibrium, it must therefore be compensated exactly by the balancing force $f_\text{balance}$.

\section{Ideal force}\label{sec:idealforce}
We can express the chemical potential as the sum of an ideal and an excess contribution, $\mu = \mu^\text{id} + \mu^\text{ex}$.
This implies that we can also split the external force [Eq.~\eqref{eq-fullF}] into an ideal and an excess contribution,
\begin{equation}
f_\text{ext} = f_\text{ext}^\text{id} + f_\text{ext}^\text{ex}.
\end{equation}

\setcounter{figure}{-1} 
\begin{figure}[t!]
\centering
\includegraphics{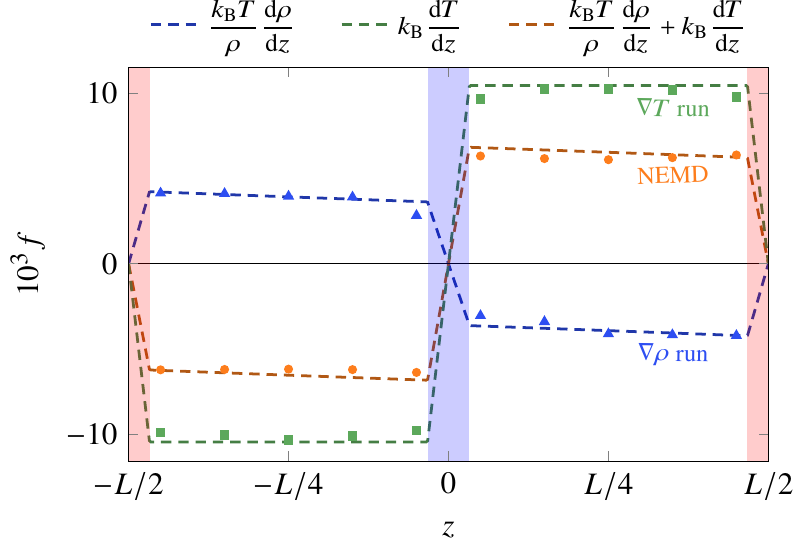}
\caption{\label{fig:ideal-force-decomposition}
Residual force measured in simulation (symbols) for the off-centre Stockmayer system with ${\redDipole}^2=1$ alongside the predicted ideal contributions from Eq.~\eqref{eq:fideval} (dashed lines). Symbols are larger than error bars.
}
\end{figure}

Our goal in this section is to relate the ideal force $f_\text{ext}^\text{id}$ to the average force experienced by
a particle in simulation,
\begin{equation}
\label{eq:fsim}
\langle f_i \rangle = \Big\langle  f_\text{ext} + \sum_j f_{ij} \Big\rangle,
\end{equation}
where $f_{ij}$ is the pairwise force computed from Eq.~\eqref{eq:phi} that is exerted on particle $i$ by particle $j$.
Because $\mu^\text{id}$ is derived from the kinetic term in the canonical partition function,
and therefore also applies to an ideal gas, we cannot interpret $f_\text{ext}^\text{id}$ as
the gradient of a potential energy. Therefore, $\langle  \sum_j f_{ij} \rangle$
will only be able to balance the \textit{excess} contribution of the external force, and
\begin{equation}
\label{eq:fsimeval}
\langle f_i \rangle = f_\text{ext}^\text{id} =  \pdc{\mu^\text{id}}{\rho}{T} \deriv{\rho}{z}  + \Biggl[s^\text{id} + \pdc{\mu^\text{id}}{T}{\rho}\Biggr] \deriv{T}{z}.
\end{equation}
We can straightforwardly evaluate the right-hand side of the above equation using the ideal chemical potential $\mu^\text{id} = k_\text{B} T \ln(\rho \Lambda^3)$
and the Sackur--Tetrode expression for the ideal entropy,
\begin{equation}
s^\text{id} =  \frac{5k_\text{B}}{2}- k_\text{B}\ln(\rho \Lambda^3),
\end{equation}
where $\Lambda$ is the de Broglie thermal wavelength.
The ideal force thus evaluates to
\begin{equation}
 \label{eq:fideval}
 f_\text{ext}^\text{id} = \frac{k_\text{B}T}{\rho} \deriv{\rho}{z} + k_\text{B} \deriv{T}{z}.
\end{equation}
In $\nabla\rho$ runs, the second term is zero, whilst in $\nabla T$ runs, the first term is zero, giving the result we used in the main text.
We note that, although our particles have orientational degrees of freedom, contributions to the ideal force due to ideal rotational motion evaluate to zero.
Furthermore, $\langle f_i \rangle$ being non-zero does not imply a continual net acceleration of the particle, because this force is balanced by the ideal pressure $P^\text{id} = \rho k_\text{B} T$ [Eq.~\eqref{eq-fullF}].

To test Eq.~\eqref{eq:fsimeval} numerically, we sampled $\langle f_i \rangle$ during production runs and compared it to the analytical expression [Eq.~\eqref{eq:fideval}], as shown in Fig.~\ref{fig:ideal-force-decomposition}.
The simulation results are in excellent agreement with the theoretical expression.

\section{Supplemental figures}\label{sec:addres}

\begin{figure}[h!]
 \centering
\includegraphics{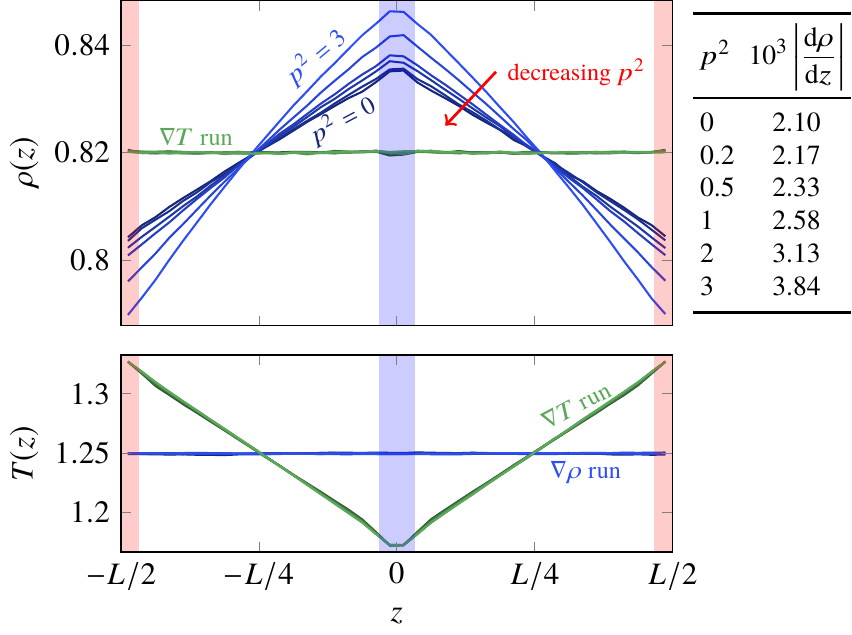}
\caption{Density and temperature gradients for the off-centre Stockmayer system of Fig.~\ref{fig:comb-costheta-stock}.
The profiles become slightly non-linear for large dipole moments. The temperature gradients do not change as a function of $\redDipole^2$; the mean gradient is $|\deriv{T}{z}|=0.010$ for all $\nabla T$ runs. However, the density gradients change significantly, and mean values of the absolute density gradients are shown in the table.}\label{fig:comb-rhoT-stock}
\end{figure}

Here, we present some additional results in support of Fig.~\ref{fig:comb-costheta-stock}.
Figure~\ref{fig:comb-rhoT-stock} provides the density and temperature profiles for the states shown in Fig.~\ref{fig:comb-costheta-stock}.
We fixed the temperatures of the hot and cold slabs to match the temperature profiles in all NEMD and $\nabla T$ runs.
The resulting density gradient increases with dipole strength.

\begin{figure}[h!]
\centering
\includegraphics{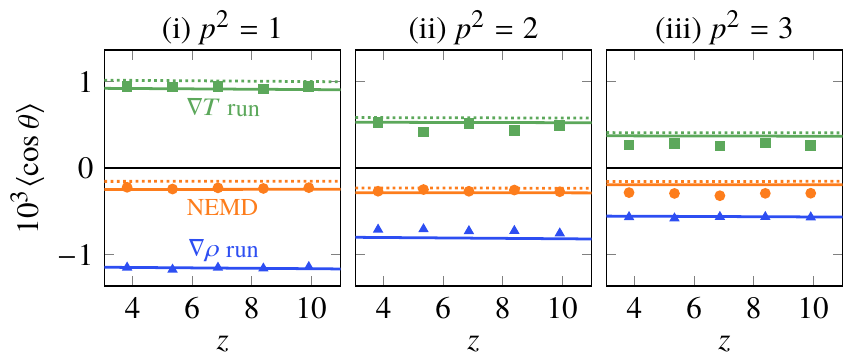}
\caption{\label{fig:comb-costheta-stock-altIdeal}
Mean orientation for an off-centre Stockmayer liquid for an alternative choice of ideal attachment sites.
Symbols represent simulation results and solid lines correspond to the theoretical predictions discussed in the main text.
Dotted lines give an alternative set of theoretical predictions where the ideal balancing force acts entirely on the LJ site for the $\nabla T$ runs.}
\end{figure}

As outlined in the main text, for $p^2 \ge 1$, we assigned half the temperature-gradient-induced ideal force to the centre of mass.
To highlight the consequences of this physically motivated but to some extent arbitrary choice,
we also present results for an alternative treatment in which the entire force acts on the LJ site (Fig.~\ref{fig:comb-costheta-stock-altIdeal}).
The ideal force only has a small effect on the results for $p^2 \ge 1$ (dotted lines), but splitting the force across the LJ and centre of mass sites yields better agreement with the simulation data (solid lines).

\end{document}